\documentclass[showpacs,preprintnumbers,twocolumn,amsmath,amssymb]{revtex4-1}
\usepackage{graphicx}
\usepackage{dcolumn}
\usepackage{bm}
\usepackage{epsfig}

\newcommand{\dhd}{{\textstyle d}
\lower.03ex\hbox{\kern-0.40em$^{\scriptstyle-}$}\kern-0.08em{}}  

\newcommand{\half}{{1\over 2}}

\newcommand{\cale}{{\cal E}}

\newcommand{\vecx}{\vec{x}}
\newcommand{\vecy}{\vec{y}}
\newcommand{\vecz}{\vec{z}}

\begin{document}

\preprint{JLAB-THY-11-1307}

\title{Mellin representation of the graviton bulk-to-bulk propagator in AdS.}

\author{Ian Baitsky }
\affiliation{
Physics Dept., Old Dominion University, Norfolk VA 23529,\\ and\\
Theory Group, Jlab, 12000 Jefferson Ave, Newport News, VA 23606}
\email{balitsky@jlab.org}

\date{\today}

\begin{abstract}
A Mellin-type representation of the graviton bulk-to-bulk propagator from Ref. \cite{gravprop} in terms of the integral over the product
of bulk-to-boundary propagators is derived.

\end{abstract}

\pacs{11.25 Tq,  11.25 Hf}

\keywords{AdS correspondence, Graviton propagator; Mellin transformation}

\maketitle

The correlation functions of the conformal ${\cal N}=4$ SYM at large coupling constant are reduced via AdS/CFT correspondence \cite{ads1,ads2,ads3} to 
Witten diagrams \cite{ads3} in AdS space.
A powerful method to calculate Witten diagrams in AdS space is to  represent  bulk-to-bulk propagators as Mellin integrals
over the bulk-to-boundary propagators, calculate the tree-level ``star'' integrals with vertices over the AdS space and then convert the remaining integrals over the flat space into Mellin 
transforms of the conformal ratios using Symanzik's star formula \cite{symanzik} (see the discussion in Ref. \cite{penedones}).  
For the scalar propagator with mass $m^2=(\Delta-d)\Delta$ the Mellin representation of bulk-to-bulk propagator has the form \cite{penedones,rizzo}
\begin{eqnarray}
&&\hspace{-2mm}
\Pi_d^\Delta(x,y)~
\label{scalprop1}\\
&&\hspace{-2mm}=~-{i\Gamma\big({d\over 2}\big)\over 4\pi^{{d\over 2}+1} }
\!\int_{-i\infty}^{i\infty}\! {d\lambda\over (\Delta-{d\over 2}\big)^2-\lambda^2}\!\!\int\! {d^dz\over\pi^{d/2}}
\nonumber\\
&&\hspace{-2mm}
\times~
{(x^0)^{{d\over 2}+\lambda}\Gamma\big({d\over 2}+\lambda\big)\over \Gamma(\lambda) [(x^0)^2+(\vecx-\vecz)^2]^{{d\over 2}+\lambda}}
{(y^0)^{{d\over 2}-\lambda}\Gamma\big({d\over 2}-\lambda\big)\over \Gamma(-\lambda)[(y^0)^2+(\vecy-\vecz)^2]^{{d\over 2}-\lambda}}
\nonumber
\end{eqnarray}
Here we used Poincare coordinates $x=(x^0,x^i)$ where $\vec{x}$ is a d-dimensional Euclidean vector (our metric is $dx^2={1\over (x^0)^2}[(dx^0)^2+d{\vec x}^2]$ 
with the size of AdS space $R=1$).
The above equation looks like the integral of the product of two bulk-to-boundary propagators with unphysical complex masses $m=\pm i\sqrt{{d^2\over 4}-\lambda^2}$ over the usual flat space and over $\lambda$. The easiest way to prove this formula is to calculate explicitly the integral over $z$ in the r.h.s. of Eq. (\ref{scalprop1}). One obtains 
(cf. Ref. \cite{ruhal})
\begin{eqnarray}
&&\hspace{-1mm} 
\Gamma\big({d\over 2}\big)\!\!\int\! {d^dz\over\pi^{d/2}}
{(x^0)^{{d\over 2}+\lambda}\Gamma\big({d\over 2}+\lambda\big)\over\Gamma(\lambda) (|x-z|^2)^{{d\over 2}+\lambda}}
{(y^0)^{{d\over 2}-\lambda}\Gamma\big({d\over 2}-\lambda\big)\over \Gamma(-\lambda)(|y-z|^2)^{{d\over 2}-\lambda}}
\nonumber\\
&&\hspace{-1mm}
=~f_\lambda(u)+f_{-\lambda}(u)
\label{integral1}
\end{eqnarray}
where $|x-z|^2\equiv (x^0)^2+(\vec{x}-\vec{z})^2$,
$$
u(x,y)={(x^0-y^0)^2+(\vec{x}-\vec{y})^2\over 2x^0y^0}
$$ 
is the chordal distance between points $x$ and $y$ and 
\begin{eqnarray}
&&\hspace{-1mm}   
f_\lambda(u)~
\label{fc}\\
&&\hspace{-1mm}  
=~
 r^{{d\over 4}+{\lambda\over 2}}(1-r)^{-{d\over 2}}
 {\Gamma\big({d\over 2}+\lambda\big)\over\Gamma(\lambda)}
F\big({d\over 2},1-{d\over 2}, 1+\lambda,{-r\over 1-r}\big)
\nonumber
\end{eqnarray}
Here $F$ is the hypergeometric function $_2F_1$ and the variable $r(u)$ is defined as
\begin{equation}
r(u)~\equiv~{1+u-\sqrt{u(2+u)}\over 1+u+\sqrt{u(2+u)}}
\label{er}
\end{equation}
Substituting the integral (\ref{integral1}) to Eq. (\ref{scalprop1}) we get
\begin{equation}
\hspace{0mm}   
\Pi_d^\Delta(u)=~-{i\over 2\pi^{{d\over 2}+1} }
\!\int_{-i\infty}^{i\infty}\! {d\lambda\over (\Delta-{d\over 2}\big)^2-\lambda^2}f_\lambda(u)
\label{int4pi}
\end{equation}
Since $r<1$ and the function 
\begin{eqnarray}
&&\hspace{-1mm} 
F\big({d\over 2},1-{d\over 2}, 1+\lambda,{-r\over 1-r}\big)~
\label{yavid}\\
&&\hspace{-1mm}=~{\Gamma(1+\lambda)\Gamma^{-1}\big({d\over 2}\big)\over\Gamma\big(1+\lambda-{d\over 2}\big)}\!\int_0^1\!dt~ (1-t)^{\lambda-{d\over 2}}\Big[t+{t^2r\over 1-r}\Big]^{{d\over 2}-1}
\nonumber
\end{eqnarray}
is regular in the right half-plane and behaves like $\lambda^{{d\over 2}-1}$ as $\Re \lambda\rightarrow \infty$ one can close the contour over $\lambda$ in Eq. (\ref{int4pi}) in the right semi-plane
and get the result as a residue at $\lambda=\Delta-{d\over 2}$ 
\begin{eqnarray}
&&\hspace{-1mm}    
\Pi_\Delta^d(u)~=~{f_{\Delta-{d\over 2}}(u)\over \pi^{d/2}(2\Delta-d)}~=~{\pi^{-{d\over 2}}\Gamma(\Delta)\over 2\Gamma\big(\Delta-{d\over 2}+1\big)}
\nonumber\\
&&\hspace{-1mm}
\times~r^{{\Delta\over 2}}(1-r)^{-{d\over 2}}
F\big({d\over 2},1-{d\over 2},\Delta-{d\over 2}+1,{-r\over 1-r}\big)
\label{Pi}
\end{eqnarray}
It is easy to see that the r.h.s. of Eq. (\ref{Pi}) is equal to the bulk-to-bulk scalar propagator \cite{scalprop}.

As we mentioned above, the formula (\ref{scalprop1}) is extremely convenient for the calculation of Witten diagrams in the Mellin representation
so it would be advantageous to get similar expression for the bulk-to-bulk graviton propagator.  This propagator can be represented as \cite{gravprop}
\begin{eqnarray}
&&\hspace{-1mm}
G^{\alpha\beta;\mu\nu}(x,y)~
\nonumber\\
&&\hspace{-1mm} =~(\partial^\alpha\partial^\mu u\partial^\beta\partial^\nu u+\alpha\leftrightarrow\beta)\Pi_d^d(u)
+g_{\alpha\beta}g_{\mu\nu}H(u)
\nonumber\\
&&\hspace{-1mm}
+\{(D^\alpha[\partial^\beta\partial^\mu u\partial^\nu u X(u)]
+D^\alpha[\partial^\beta u\partial^\mu u\partial^\nu u Y(u)]
\nonumber\\
&&\hspace{-1mm} 
+~\alpha\leftrightarrow\beta)~+~D^\alpha[\partial^\beta Z(u)]g^{\mu\nu}+(\alpha\leftrightarrow\mu,\beta\leftrightarrow\nu)\}
\label{gravprop}
\end{eqnarray}
where $D_\mu$ is a covariant derivative and 
\begin{eqnarray}
&&\hspace{-0mm}
H(u)~=~-{2\over d-1}\Big[(1+u)^2 \Pi_{d}^d(u)
\nonumber\\
&&\hspace{20mm}
~-(d-2)(1+u)\!\int_u^\infty\!du'~\Pi_d^d(u')\Big]
\label{H}
\end{eqnarray}
The remaining three functions $X(u)$, $Y(u)$ and $Z(u)$ are gauge artifacts. Hereafter
the Greek indices from the first half of alphabet refer to the point $x$ and from the second to $y$.

The Mellin representation of the graviton propagator has the form 
\begin{eqnarray}
&&\hspace{-1mm}
G^{\alpha\beta;\mu\nu}(x,y)~
\label{mygravprop}\\
&&\hspace{-1mm}
=~{i\Gamma(d/2)\over 2(d-1)\pi^{{d\over 2}+1} }
\!\int_{-i\infty}^{i\infty}\! {d\lambda\over (d/2)^2-\lambda^2}
{\big({d\over 2}+1\big)^2-\lambda^2\over \Gamma(\lambda)\Gamma(-\lambda)}
\nonumber\\
&&\hspace{-1mm}
\!\!\int\! {d^dz\over\pi^{d/2}}
{(x_0)^{{d\over 2}+\lambda+2}\Gamma\big({d\over 2}+\lambda\big)\over (|x-z|^2)^{{d\over 2}+\lambda}}{(y^0)^{{d\over 2}-\lambda+2}\Gamma\big({d\over 2}-\lambda\big)\over (|y-z|^2)^{{d\over 2}-\lambda}}
\nonumber\\
&&\hspace{-1mm}  
\times~J^{\alpha i}(x-z)J^{\beta j}(x-z)
\cale_{ij;kl}
J^{k\mu}(z-y)J^{l\nu}(z-y)
\nonumber
\end{eqnarray}
where 
\begin{eqnarray}
&&\hspace{-1mm}
J^{\mu i}(x-z)~=~\delta^{\mu i}-2{(x-z)^\mu(x-z)^i\over |x-z|^2}
\end{eqnarray}
(and similarly for other $J'$s) while ${\cal E}_{ij;kl}$ is a traceless symmetric projector
\begin{eqnarray}
&&\hspace{-1mm}
{\cal E}_{ij;kl}~=~\half(\delta_{ik}\delta_{jl}+\delta_{il}\delta_{jk})-{\delta_{ij}\delta_{kl}\over d}
\end{eqnarray}
The $d$-dimensional Latin indices of this projector are raised and lowered  with the flat metric.

Note that the covariant derivative and the trace of the graviton propagator (\ref{mygravprop}) vanish:  
\begin{equation}
g_{\alpha\beta}G^{\alpha\beta;\mu\nu}(x,y)=0,~~~~~~~~~~~~~D_\alpha G^{\alpha\beta;\mu\nu}(x,y)=0
\label{gaugecondition}
\end{equation}
Let us compare the integrand in the formula (\ref{mygravprop}) to bulk-to-boundary propagator of the graviton. The general solution of the 
 Dirichlet problem with the boundary data $\hat{h}_{ab}$ has the form \cite{tliu}:
\begin{eqnarray}
&&\hspace{-1mm} 
h^\alpha_\beta(x)~=~{(d+1)\Gamma(d)\over (d-1)\Gamma(d/2)}
\label{bulktobou}\\
&&\hspace{-1mm}  
\times~\!\int\! {d^dz\over\pi^{d/2}}{(x^0)^d\over (|x-z|^2)^d}J^{\alpha i}(x-z)J^{\beta j}(x-z)
\cale_{ij;ab}\hat{h}_{ab}
\nonumber
 \end{eqnarray}
We see that similarly to the scalar case, the Mellin representation (\ref{mygravprop}) looks like an
 integral of the product of two bulk-to-boundary propagators with unphysical complex graviton masses 
 $m=\pm i\sqrt{{d^2\over 4}-\lambda^2}$ over the usual flat space and over $\lambda$.

Now let us prove the Eq.  (\ref{mygravprop}). The central point of the proof is the calculation of the following integral
\begin{eqnarray} 
 &&\hspace{-1mm}
 I^{\alpha\beta;\mu\nu}(x,y;\lambda)~=~2\big[\big({d\over 2}+1\big)^2-\lambda^2\big]\Gamma(d/2)
 \label{master1}\\
&&\hspace{-1mm}  
\times \int\! {d^dz\over\pi^{d/2}}
{(x^0)^{{d\over 2}+\lambda-2}\Gamma\big({d\over 2}+\lambda\big)\over \Gamma(\lambda)(|x-z|^2)^{{d\over 2}+\lambda}}
{(y^0)^{{d\over 2}-\lambda-2}\Gamma\big({d\over 2}-\lambda\big)\over \Gamma(-\lambda)(|y-z|^2)^{{d\over 2}-\lambda}}
\nonumber\\
&&\hspace{-1mm}  
\times~J^{\alpha i}(x-z)J^{\beta j}(x-z)
\cale_{ij;kl}
J^{k\mu}(z-y)J^{l\nu}(z-y)
 \nonumber
\end{eqnarray}
It can be decomposed in the same set of structures as the propagator (\ref{gravprop})
\begin{eqnarray} 
 &&\hspace{-1mm}
 I^{\alpha\beta;\mu\nu}(x,y;\lambda) 
 \nonumber\\
&&\hspace{-1mm}
=~(\partial^\alpha\partial^\mu u\partial^\beta\partial^\nu u+\alpha\leftrightarrow\beta)G_\lambda(u)+g^{\alpha\beta}g^{\mu\nu}H_\lambda(u)
\nonumber\\
&&\hspace{-1mm}
+~\{(D^\alpha[\partial^\beta\partial^\mu u\partial^\nu u X_\lambda(u)]+D^\alpha[\partial^\beta u\partial^\mu u\partial^\nu u Y_\lambda(u)]
\nonumber\\
&&\hspace{29mm}
+~\alpha\leftrightarrow\beta)+~(\alpha\leftrightarrow\mu,\beta\leftrightarrow\mu)\}
\nonumber\\
&&\hspace{-1mm} 
+~D^\alpha[\partial^\beta Z_\lambda(u)]g^{\mu\nu}+D^\mu[\partial^\nu Z_{-\lambda}(u)]g^{\alpha\beta}
\label{master2}
\end{eqnarray}
 A straightforward but somewhat lenghtly calculation yields (cf. Ref. \cite{ruhal})
\begin{eqnarray} 
 \hspace{-5mm}
 G_\lambda(u)~&=&~\big[\big({d\over 2}-1\big)^2-\lambda^2\big]f_\lambda(u)~+~(\lambda\leftrightarrow -\lambda)
 \nonumber\\
\hspace{-5mm}
H_\lambda(u)~&=&~2(1+u)^2f_\lambda(u)-{2\over d}\big({d^2\over 4}-\lambda^2\big)f_\lambda(u)
\nonumber\\
\hspace{-5mm}
&+&~2(d-2)(1+u)F_\lambda(u)~+~(\lambda\leftrightarrow -\lambda)
\label{master3}
\end{eqnarray}
for  the two physical structures and
\begin{eqnarray} 
 &&\hspace{-1mm}
\Big[{d^2\over 4}-\lambda^2\Big]X_\lambda(u)~=~\big[(1+u)^2-{1\over d}\big]f''_\lambda(u)+~\big[\big({d\over 2}+1\big)^2
 \nonumber\\
&&\hspace{-1mm}
-~\lambda^2\big](1+u)f'_\lambda(u)+d\big({d^2\over 4}-\lambda^2\big)f_\lambda(u~+~(\lambda\leftrightarrow -\lambda)),
 \nonumber\\
 &&\hspace{-1mm}
\Big[{d^2\over 4}-\lambda^2\big]Y_\lambda(u)~=~\Big[(1+u)^2-{1\over d}\big]f'''_\lambda(u)
 \nonumber\\
&&\hspace{-1mm}
+~(d+1)(1+u)f''_\lambda(u)+{d(d+1)\over 2}f'_\lambda(u) ~+~(\lambda\leftrightarrow -\lambda),
 \nonumber\\
 &&\hspace{-1mm}
\Big[{d^2\over 4}-\lambda^2\big]Z_\lambda(u)~=~\Big[(1+u)^3-{1\over d}\big][f''_\lambda(u)+f''_{-\lambda}(u)]
\nonumber\\
&&\hspace{-1mm}
+~\Big[(1+d-2\lambda)(1+u)+\big({d\over 2}+{2\over d}\lambda^2-{1\over d}\big)\Big]
\nonumber\\
&&\hspace{-1mm}
\times~[f'_\lambda(u)+f'_{-\lambda}(u)]~+~2(d-1)\lambda(1+u)[f_\lambda(u)+f_{-\lambda}(u)]
\nonumber\\
&&\hspace{-1mm}
+~\big[2(d-1)\lambda+(2-d-2\lambda)\big({d^2\over 4}-\lambda^2\big)\big][F_\lambda(u)+F_{-\lambda}(u)]
\nonumber\\
\label{master4}
\end{eqnarray}
for  three gauge-dependent ones.  Here 
\begin{eqnarray} 
&&\hspace{-1mm}
F_\lambda(u)~=~-\int_u^\infty f_\lambda(v)dv~=~-{\Gamma\big({d\over 2}+\lambda\big)\over\Gamma(d/2)}
\label{Fc}\\
&&\hspace{-1mm}
\times~{r^{{\lambda\over 2}+{d-2\over 4}}(1-r)^{1-{d\over2}}\over \big(d-2+2\lambda\big)}F\big({d\over 2}-1,2-{d\over 2},1+\lambda,{-r\over 1-r}\big)
\nonumber
\end{eqnarray}
One can easily see that the function $F\big({d\over 2}-1,2-{d\over 2},1+\lambda,{-r\over 1-r}\big)$ is also regular at the right half-plane and behaves
like $\lambda^{{d\over 2}-1}$ as $\Re \lambda\rightarrow\infty$, cf. Eq. (\ref{yavid}).

Let us now return to the proof of Eq. (\ref{mygravprop}) which can be rewritten as
\begin{eqnarray}
&&\hspace{-1mm}
G^{\alpha\beta;\mu\nu}(x,y)~
\label{mygravprop1}\\
&&\hspace{-1mm}
=~{i(d-1)^{-1}\over 4\pi^{{d\over 2}+1} }
\!\int_{-i\infty}^{i\infty}\! {d\lambda\over (d/2)^2-\lambda^2}I^{\alpha\beta;\mu\nu}(x,y,\lambda)
\nonumber
\end{eqnarray}
Let us discuss the two gauge-invariant structures $G(u)$ and $H(u)$. The corresponding terms in the r.h.s of Eq. (\ref{mygravprop1}) are
\begin{eqnarray}
&&\hspace{-0mm}
(\partial^\alpha\partial^\mu u\partial^\beta\partial^\nu u+\alpha\leftrightarrow\beta)
{i(d-1)^{-1}\over 4\pi^{{d\over 2}+1} }
\!\int_{-i\infty}^{i\infty}\! {d\lambda\over {d^2\over 4}-\lambda^2}
G_\lambda(u)
\nonumber\\
&&\hspace{-0mm}
+~g^{\alpha\beta}g^{\mu\nu}{i(d-1)^{-1}\over 4\pi^{{d\over 2}+1} }
\!\int_{-i\infty}^{i\infty}\! {d\lambda\over {d^2\over 4}-\lambda^2}
H_\lambda(u)
\nonumber\\
&&\hspace{-0mm}
=~(\partial^\alpha\partial^\mu u\partial^\beta\partial^\nu u+\alpha\leftrightarrow\beta)
{i(d-1)^{-1}\over 2\pi^{{d\over 2}+1} }
\!\int_{-i\infty}^{i\infty}\! {d\lambda\over {d^2\over 4}-\lambda^2}
\nonumber\\
&&\hspace{50mm}
\times~\big[\big({d\over 2}-1\big)^2-\lambda^2\big]f_\lambda(u)
\nonumber\\
&&\hspace{-0mm}
+~g^{\alpha\beta}g^{\mu\nu}{i(d-1)^{-1}\over 2\pi^{{d\over 2}+1} }
\!\int_{-i\infty}^{i\infty}\! {d\lambda\over {d^2\over 4}-\lambda^2}[2(1+u)^2f_\lambda(u)
\nonumber\\
&&\hspace{11mm}
-~{2\over d}\big({d^2\over 4}-\lambda^2\big)f_\lambda(u)+2(d-2)(1+u)F_\lambda(u)]
\end{eqnarray}
As we discussed above (see Eqs. (\ref{fc}), (\ref{yavid}), and (\ref{Fc})), the functions $f_\lambda(u)$ and $F_\lambda(u)$ 
are regular in the right half-plane and decrease as $\lambda^{{d\over 2}-1}e^{-{\lambda\over 2}|\ln r|}$ when $\Re \lambda\rightarrow\infty$  so one can close the contour over $\lambda$ and take the residue at $\lambda={d\over 2}$.
One obtains 
\begin{eqnarray}
&&\hspace{-1mm}
G^{\alpha\beta;\mu\nu}(x,y)~=~{f_{d/2}(u)\over d\pi^{d/2}}(\partial^\alpha\partial^\mu u\partial^\beta\partial^\nu u+\alpha\leftrightarrow\beta)
\nonumber\\
&&\hspace{-1mm}
-~{2\over(d-1)\pi^{d/2}}\big[(1+u)^2f_{d\over 2}(u)+(d-2)(1+u)F_{d\over 2}(u)\big]
\nonumber\\
&&\hspace{-1mm}
\times~g^{\alpha\beta}g^{\mu\nu}~+~{\rm gauge-dependent~structures}
\label{mygravprop3}
\end{eqnarray}
which coincides with Eq. (\ref{gravprop}) and Eq. (\ref{H}) since $\Pi_d^d(u)~=~{1\over d\pi^{d/2}}f_{{d\over 2}}(u)$. Thus, we proved that the integral (\ref{mygravprop}) 
can serve as a gravition bulk-to-bulk propagator in the gauge $D_\alpha G^{\alpha\beta;\mu\nu}~=~0$. It should be mentioned that similar but somewhat more complicated representation of the graviton
propagator was obtained in Ref. \cite{rizzo}). It has a function ${({d\over 2}+1)^2-\lambda^2\over ({d\over 2}-1)^2-\lambda^2}$ in place of 
${({d\over 2}+1)^2-\lambda^2\over 1-d}$ in Eq. (\ref{mygravprop}) as well as additional terms proportional to  
the tensor structure obtained from that of Eq. (\ref{mygravprop}) by replacement $\cale_{ij;kl}~\rightarrow~\delta_{ij}\delta_{kl}$ and to the $g^{\mu\nu}g^{\alpha\beta}$ structure. 

For completeness, let us briefly discuss  the gauge boson propagator \cite{gbosonprop}
\begin{equation}
G_{\alpha;\mu}(x,y)~=~-{f_{{d\over 2}-1}(u)\over 2\pi^{d/2}\big({d\over 2}-1\big)}~\partial^\mu\partial^\alpha u+\partial^\mu\partial^\nu S(u)
\end{equation}
where the second structure depends on the choice of gauge. 
The Mellin representaton of this propagator has the form \cite{rizzo}
\begin{eqnarray}
&&\hspace{-1mm}
G^{\alpha;\mu}(x,y)~
\label{gbosonprop}\\
&&\hspace{-1mm}
=~{i\Gamma(d/2)\over 4\pi^{{d\over 2}+1} }
\!\int_{-i\infty}^{i\infty}\! d\lambda~{(d/2)^2-\lambda^2\over \big[\big({d\over 2}+1\big)^2-\lambda^2\big]^2}
\nonumber\\
&&\hspace{-1mm}
\!\!\int\! {d^dz\over\pi^{d/2}}
{(x_0)^{{d\over 2}+\lambda+2}\Gamma\big({d\over 2}+\lambda\big)\over \Gamma(\lambda)(|x-z|^2)^{{d\over 2}+\lambda}}{(y^0)^{{d\over 2}-\lambda+2}\Gamma\big({d\over 2}-\lambda\big)\over \Gamma(-\lambda)(|y-z|^2)^{{d\over 2}-\lambda}}
\nonumber\\
&&\hspace{-1mm}  
\times~J^{\alpha i}(x-z)
\delta_{ik}
J^{k\mu}(z-y)
\nonumber
\end{eqnarray}
The explicit calculation of the integral in the r.h.s. of this equation confirms this expression obtained in Ref. \cite{rizzo}
by solution of Einstein equations.
Again, the gauge condition for the propagator (\ref{gbosonprop}) is $D_\alpha G^{\alpha;\mu}(x,y)~=~0$.

We have represented the graviton bulk-to-bulk propagator in the form of the Mellin integral of the product of bulk-to-boundary
propagators (with nonphysical masses). This formula permits us to apply the Mellin-transformation method of Ref. \cite{penedones}  to Witten diagrams  with graviton (and gauge boson)
propagators. 

\section*{Acknowledgements}
The author is grateful to J. Penedones for valuable discussions.
This work was supported by contract
 DE-AC05-06OR23177 under which the Jefferson Science Associates, LLC operate the Thomas Jefferson National Accelerator Facility. 
 
 
\end{document}